%% file: main.tex
\documentclass[sigchi]{acmart}

\usepackage{multirow}
\usepackage{graphicx}
\usepackage{enumerate}
\AtBeginDocument{%
  \providecommand\BibTeX{{%
    \normalfont B\kern-0.5em{\scshape i\kern-0.25em b}\kern-0.8em\TeX}}}

\copyrightyear{2022}
\acmYear{2022}
\setcopyright{rightsretained}
\acmConference[ICTD2022]{International Conference on Information \& Communication Technologies and Development 2022}{June 27--29, 2022}{Seattle, WA, USA}
\acmBooktitle{International Conference on Information \& Communication Technologies and Development 2022 (ICTD2022), June 27--29, 2022, Seattle, WA, USA}
\acmDOI{10.1145/3572334.3572372}
\acmISBN{978-1-4503-9787-2/22/06}
\begin{document}

\title{Tackling Hate Speech in Low-resource Languages \\with Context Experts}

\author{Daniel Nkemelu}
\email{dnkemelu@gatech.edu}
\affiliation{%
  \institution{Georgia Institute of Technology}
  \city{Atlanta}
  \state{Georgia}
  \country{USA}
}

\author{Harshil Shah}
\email{harshil18@gatech.edu}
\affiliation{%
  \institution{Georgia Institute of Technology}
  \city{Atlanta}
  \state{Georgia}
  \country{USA}
}

\author{Irfan Essa}
\email{irfan@gatech.edu}
\affiliation{%
  \institution{Georgia Institute of Technology}
  \city{Atlanta}
  \state{Georgia}
  \country{USA}
}

\author{Michael L. Best}
\email{mikeb@gatech.edu}
\affiliation{%
  \institution{Georgia Institute of Technology}
  \city{Atlanta}
  \state{Georgia}
  \country{USA}
}

\renewcommand{\shortauthors}{Nkemelu et al.}

\begin{abstract}
  Given Myanmar's historical and socio-political context, hate speech spread on social media have escalated into offline unrest and violence. This paper presents findings from our remote study on the automatic detection of hate speech online in Myanmar. We argue that effectively addressing this problem will require community-based approaches that combine the knowledge of context experts with machine learning tools that can analyze the vast amount of data produced. To this end, we develop a systematic process to facilitate this collaboration covering key aspects of data collection, annotation, and model validation strategies. We highlight challenges in this area stemming from small and imbalanced datasets, the need to balance non-glamorous data work and stakeholder priorities, and closed data-sharing practices. Stemming from these findings, we discuss avenues for further work in developing and deploying hate speech detection systems for low-resource languages.
\end{abstract}



\maketitle

\section{Introduction}
\input{sections/introduction.tex}

\section{Background}
\input{sections/relatedwork.tex}

\section{Automatic Hate Speech Detection with Context Experts}
\input{sections/methodology.tex}

\section{Study findings \& Discussion}
\input{sections/findings.tex}

\section{Conclusion}
\input{sections/conclusion.tex}

\section{Future Work}
\input{sections/future.tex}

\begin{acks}
Special thanks to our partners at The Carter Center for collaborating with us in Myanmar and establishing our partnership with the New Myanmar Foundation (NMF); to the NMF for working with us and serving as context experts despite challenges with the COVID-19 pandemic; and to Koe Koe Tech for their review and feedback on an initial version of the hate speech annotation guideline.
\end{acks}

\bibliographystyle{ACM-Reference-Format}
\bibliography{main}

\appendix
\section{Appendix}
This Github link contains data on the Burmese hate lexicons (Hatebase.org and Phandeeyar), adopted protected characteristics, annotation guideline and plan: \url{https://github.com/TID-Lab/myanmarhsc}.

\end{document}

%% file: sections/introduction.tex
The rapid adoption of social media in Myanmar has been accompanied by a surge in the dissemination of problematic content such as hate speech, disinformation, and misinformation~\cite{kyaw2019facebooking}. While platforms like Facebook offer opportunities for people to connect online,  do business, and participate in online activism~\cite{best2016mobile}, they have also seen use as a medium for inciting violence against minority groups from online actors~\cite{mclaughlin2018facebook, muller2021fanning}. In 2018, a United Nations independent fact-finding report highlighted the role social media, specifically Facebook, played in spreading hate speech and disinformation that led to a genocide of the Rohingya ethnic minority group in Myanmar. Lee~\cite{lee2019extreme} discussed the role of citizen-generated posts and state media-led publication outlets in spreading anti-minority rhetoric that influenced violent narratives about the Rohingya on social media. Violence towards the Rohingya community spurred by the proliferation of hate speech and disinformation on Facebook~\cite{fink2018dangerous} led to the murder of thousands of civilians, creating almost a million refugees~\cite{un2018report, warofka2018independent}. Myanmar presents a chilling yet increasingly familiar account of the weaponization of Facebook in a nation with a history of armed conflict, authoritarianism, and censorship.

Outside Myanmar, similar challenges with hate speech have been witnessed in countries like Ethiopia~\cite{gagliardone2016mechachal} and Sri Lanka~\cite{wijeratne2018control}. There is a dearth of work investigating effective strategies for real-world hate speech detection in low-resource languages. Current strategies for tackling hate speech in low-resource contexts primarily entail two main steps: exploring the use of sophisticated machine learning tools for detecting hate speech and contracting human content moderators to flag, demote, and ultimately remove problematic content. Both approaches present notable limitations. Machine learning systems require ground truth data and data processing capabilities that are not readily available with low-resource languages. In addition, the vast amount of social media content produced daily makes it infeasible to engage human trackers to detect every instance of hate speech except the most prominent. Even if this was possible for a single context, it is not trivial to scale to new contexts, languages, and countries. 

We seek to address this need by exploring a community-driven approach to tackle hate speech in low-resource language settings. This approach involves working with \textit{context experts} on the entire machine learning project pipeline: scoping the project with a local partner focused on issues related to digital threats to democracy, assessing hate speech definitions and guidelines in tandem with legal experts, and working with paid volunteers to generate quality data, train, and validate machine learning models. We use the term \textit{context experts} to highlight their role not merely as language translators but as experts with deep and personal knowledge of the context resulting from their lived experience. We interchangeably use the terms annotator and context experts in some parts of this paper. We only refer to the context experts as annotators when maintaining standard language for data labeling tasks. 

This paper offers a report on our remote study of hate speech in Myanmar. In the months leading up to the 2020 Myanmar national elections, we worked remotely, due to the COVID-19 pandemic, with context experts to curate a dataset of 226 Burmese hate speech posts from Facebook through its CrowdTangle API service~\cite{team2020crowdtangle}. Our work contributes to research in machine learning for development (ML4D) seeking to understand ways to tackle hate speech on social media. We develop a process for coordinating machine learning work within low-resource language settings and show that working with context experts offers a key solution to the problem of hate speech in these settings. We also provide an early look into downstream classification tasks for the Burmese language using classical and neural network-based machine learning models.

%% file: sections/relatedwork.tex
This section provides some background on Myanmar's political history and the evolution of hate speech targeted explicitly against the Rohingya community. We broadly discuss hate speech detection algorithms and datasets in resource-rich and low-resource languages and highlight Burmese natural language processing work.

\subsection{Social Media in Myanmar and Hate Speech}
After gaining independence in January 1948, Myanmar's over 14 years-long democratic government was interrupted by a military coup and subsequent dictatorship in 1962 that lasted almost 50 years. The military at the time pushed against calls for autonomy by non-Burman ethnic groups, which they labeled as anti-nationalist and anti-unity. For most of its years post-independence, Myanmar has faced many ethnic and religious conflicts and wars. When Myanmar re-transitioned to democratic rule in 2011, these conflicts remained persistent~\cite{zin2015anti}. The government was dominated by the majority Bamar Buddhists who had exclusive control of military and civilian institutions, despite the country's cultural and linguistic diversity. This unequal power share led to the marginalization of ethnic minority groups resulting in both armed conflict and non-violent political actions~\cite{smith1991burma, thawnghmung2011beyond}. 

An estimated 1 million Rohingyas living in the Rakhine state have historically faced discriminatory practices from the military government, including but not limited to restricted access to education, employment, and citizenship identity cards~\cite{zin2015anti, fink2018dangerous}. These practices appeared to have worsened since the transition to democracy. Lee~\cite{lee2016dark} pointed to the liberalization of media and political freedom stemming from the transition as a motivating factor that amplified political polarization, which fuelled the agendas of ultra-nationalist anti-Rohingya Buddhist groups.

McLaughlin~\cite{mclaughlin2018facebook} discusses how violence against the Rohingya was preceded by the viral spread of hateful rhetoric and disinformation on Facebook primarily targeted against Muslims. Facebook experienced massive growth in adoption between 2016 and 2017 through its ``free basics'' program, which allowed users to sign up for a free, limited Facebook version without a mobile internet plan~\cite{best2014internet, mclaughlin2018facebook}. According to Fink ~\cite{fink2018dangerous}, Facebook ignored several warnings by local rights and technology civil society organizations to act on dangerous speech posted on the platform calling on Buddhists to pick up arms in preparation for Muslim attacks and vice versa in 2017. The United Nation's independent fact-finding mission on Myanmar confirmed the significant negative role of hate speech and disinformation spread on Facebook played in heating the polity~\cite{un2018report}. The platform has struggled to effectively moderate content in Myanmar's context due to the non-locally resident nature of its moderation system. The escalation of conflict in August 2017 lasted more than two months and led to over 750,000 fleeing as refugees~\cite{un2018report}. 

As Myanmar geared for its national election in late 2020, observers were concerned about the possibility of online actors exploiting existing distrust of the government and media in the country to foster violent responses to election results. This worry prompted opportunities for local and international civil society organizations to explore localized hate speech tracking and mitigation projects. It is within this socio-political context that this work is situated.

\subsection{Automated Hate Speech Detection}
Earlier works in hate speech detection have mostly leveraged a keyword-based approach that relied on the presence of a derogatory term typically used in hate speech to make decisions about whether a post is hate speech or not, e.g.~\cite{kontostathis2013detecting}. However, keyword-based approaches have been shown to offer little performance value \cite{saleem2017web, macavaney2019hate}. More sophisticated computational approaches for tackling online hate speech have gained attention in recent years, and machine learning techniques have since been applied to hate speech detection~\cite{schmidt2017survey, waseem2016you, saleem2017web}. Prior work have explored bag-of-words, word-, and character-level n-grams features~\cite{mehdad2016characters, nobata2016abusive, waseem2016hateful} and TF/IDF weighted embedding methods~\cite{davidson2017automated}, with algorithms such as support vector machines~\cite{malmasi2018challenges}, balanced random forests\cite{burnap2015cyber}, and logistic regression models~\cite{davidson2017automated}. Recent works have adopted neural network-type approaches such as convolutional neural networks (CNNs)~\cite{gamback2017using}, CNNs combined with a Gated Recurrent Unit (GRU) network~\cite{zhang2018detecting}, and Transformer-based models such as the Bidirectional Encoder Representations from Transformers (BERT)~\cite{mozafari2019bert, mozafari2020hate}.

Data is a critical resource for determining performance, robustness, and scalability in machine learning systems~\cite{halevy2009unreasonable}. To this end, hate speech datasets have been released by researchers in this area~\cite{burnap2016us, waseem2016hateful, waseem2016you, davidson2017automated, founta2018large}. In their survey of the hate speech detection literature, Fortuna and Nunes~\cite{fortuna2018survey} found the majority of the datasets to be in English, with few exceptions in Dutch~\cite{tulkens2016dictionary}, German~\cite{ross2017measuring}, and Italian~\cite{del2017hate}. These datasets range in size from as small as 36 tweets~\cite{benikova2017does} to as large as 150,000 multimodel (image-text) tweets~\cite{gomez2020exploring} sourced from publicly available internet data. 

A large proportion of the works mentioned earlier (e.g., ~\cite{chatzakou2017mean, founta2018large}) leverage crowdsourcing platforms like Figure-Eight (formerly Crowdflower)~\cite{appen2021} or use publicly sourced comments online as ground truth (e.g., Warner and Hirschberg~\cite{warner2012detecting} used several thousand comments from Yahoo!). In some cases, the research team labels the data themselves. We note that several of these resources may not be available in many low-resource contexts. First, crowdsourcing platforms do not have universal coverage across languages and geographic regions. Even if they did, finding the right annotators for a hate speech task that requires knowledge of social contexts can be challenging. There is also not a broad diversity of web platforms in low-resource language settings to scrape potential ground truth data. Most times, people who understand the language and the socio-political context may not be the researchers themselves, thus requiring the kind of collaboration we explore in this paper. 

\subsubsection{Low-Resource Hate Speech Detection Data}
We refer to low-resource strictly within the context of limited availability of technical resources such as labeled training data; linguistic tools for tasks such as semantic analysis, named-entity recognition, and parts of speech tagging; or digitized texts that can serve as supervised/unsupervised training data for language models. Researchers have also explored the task of hate speech detection in these contexts. Mubarak et al.~\cite{mubarak2017abusive} studied the use of abusive language in the Arabic language on Twitter. Ishmam and Sharmin~\cite{ishmam2019hateful} explored machine learning approaches for classifying public Facebook posts in the Bengali language. Similar studies have been conducted in Amharic~\cite{mossie2018social}, Indonesian~\cite{alfina2017hate}, Hindi~\cite{sreelakshmi2020detection}, and Vietnamese~\cite{van2019hate}. Our investigation into these works shows that none point to the relevance of working with context experts as central to scaling hate speech detection in low-resource contexts. The authors also offered limited visibility into the data curation process, making it difficult to replicate it in new environments. 

In their critical analysis of existing hate speech detection datasets, Madukwe et al.~\cite{madukwe2020data} highlight that several works do not make their data publicly available, making it difficult to benchmark. When provided on request, the data may suffer from data degradation---a case where a dataset re-generated on demand by the researcher no longer produces the same amount or quality of data as at the time of publication. While data collection and sharing are vital for scientific progress, we acknowledge that authors cannot often do so for several reasons, including privacy concerns, platform restrictions, and the potential dissemination of harmful content. Anane-Sarpong et al.~\cite{anane2018you} discuss how various structural, organizational, cultural, and ethical complexities influence a researcher's decision to share their data. For instance, according to CrowdTangle's terms of use, we do not have Facebook's permission to share our dataset from Myanmar. However, we detail our steps for the hate speech dataset curation and provide materials for future use in new contexts and possible replication studies. 

\subsection{Burmese Language Processing}
Burmese is the official language of Myanmar and the native language of the Bamar people. It is a largely monosyllabic and analytic language with subject-object-verb word ordering and belongs to the Sino-Tibetan family. Like Chinese, Burmese morphemes can be combined freely with no changes~\cite{jenny2017burmese}. Ding et al.~\cite{ding2016word} describes a challenge that arises for Burmese language processing because the boundaries of what implies a ``word'' are not clearly defined. This challenge emerges because Burmese has no specific rule or convention on how spaces separate words. Traditional Burmese does not use white spaces to mark word boundaries. However, modern variants of Burmese do use white spaces between phrases to improve readability. 

Another significant challenge is the lack of consistent, standardized font encoding. The most widely-used Burmese font for reading and writing on modern computers and smartphones is the Zawgyi font. Zawgyi is not defined as a standard character encoding and is not part of the standard Unicode character set. As a result, it is not typically built into major operating systems. Nonetheless, Facebook supports Zawgyi as an optional encoding on their app and website, and this option is widely used in Myanmar.

The Burmese language is referred to as part of a class of low-resource languages due to the limited availability of tools, datasets, and studies focused on the language. A series of recent works have been published, laying the groundwork for Burmese natural language processing. These include morphological analyses such as syllable-based tokenization~\cite{ding2019towards}, part-of-speech tagging~\cite{ding2018nova}, word segmentation~\cite{ding2016word}, named-entity transliteration~\cite{mon2020myanmar}, and the development of a Burmese treebank~\cite{ding2020burmese} as part of the Asian Language Treebank Project~\cite{riza2016introduction}. Burmese has also been featured as a constituent language in monolingual~\cite{jiang2021pre} and multilingual~\cite{conneau2019unsupervised} learning tasks for language models. This work provides an early look into downstream classification tasks for automatic hate speech detection in the Burmese language using classical and neural network-based machine learning models. 

%% file: sections/methodology.tex
Most automatic hate speech detection systems rely on machine learning algorithms trained on existing ground truth data. However, the resources needed to facilitate this task are limited to a small set of languages and organizations. Similarly, publicly-available pretrained large language models only work on a handful of languages. This limitation implies that progress in automatic hate speech detection tends to follow a top-down approach where only privileged languages gain engineering attention. To balance this disproportion, we aim to design a collaborative process with context experts in low-resource language settings to scope the problem of hate speech detection and develop machine learning models to address them.

To concretize this collaboration, we focus on the task of developing machine learning models in the Burmese language to automatically detect hate speech posted on social media within the context of the Myanmar general election. A process approach helps us temporally order the crucial aspects of this work to support replication, reuse, and revalidation. By developing this process, we hope to provide a recipe for practitioners and researchers interested in collaborative work that addresses hate speech in low-resource language settings. Our approach consists of the following four key steps:
\begin{enumerate}
    \item establish partnerships to co-design project scope, identify and recruit paid volunteers.
    \item contextualize hate speech definitions and annotation guidelines for local relevance with legal experts.
    \item generate quality data and train machine learning models.
    \item validate trained machine learning models, and iterate step (3).
\end{enumerate}

To facilitate this process, context experts take on two roles. First, they serve as technologists, actively contributing to shaping the overall direction of our research, and subsequently as model validators, identifying opportunities for improving the machine learning system.

\subsection{Context Experts as Technologists}
Attygale~\cite{attygalle2017context} discusses the idea of engaging with context experts as an intentional process of co-creating solutions in partnership with people who know the opportunities for and barriers to impact through their own experiences. According to Attygale~\cite{attygalle2017context}, context experts offer perspectives that add depth and breadth to the technical expertise of ``content experts'' (in our case, machine learning researchers). In this work, we adopt context experts to refer to Burmese natives working in civil society and with an expressed interest in their nation's political system. \textbf{We argue that sustainably addressing hate speech in low-resource language settings involves empowering context experts with the technology-driven tools needed to tackle the problem locally}. The goal is to center the voices of the context experts from the onset of the project and ensure they have a sense of ownership. This type of collaborative model is often broadly described as community consultation, deliberation, engagement, or participation~\cite{ranasinghe2018engaged}. It differs from simply relying on them for feedback or buy-in, as in a contextual inquiry.

\subsubsection{Recruitment and Training} 
Our partners in Myanmar advertised recruitment calls on public social and political Facebook pages. These pages were selected due to the number of accounts following the page and its focus on social and political issues. The call sought people who were active on Facebook and interested in the Myanmar's political system. The earliest 12 respondents were interviewed on a roll-in basis. The interviews were conducted by our resident project manager and focused on the interviewees' knowledge of relevant political issues and their ability to use a computer. After reviews and interviews, eight people were recruited. Five of them identified as female and three identified as male. All the context experts were native Burmese speakers and resident in Myanmar. The context experts attended an initial virtual session to connect and discuss the motivation for the project concerning the forthcoming Myanmar national elections. A subsequent training session to prepare the context experts for the data curation process necessary for the task. We provided labeling guidelines in the English and Burmese languages. The training was facilitated by an experienced project manager, an expert in Myanmar politics. The facilitator conducted the training session in the Burmese language through a video conference call that lasted for 90 minutes. We then created a shared workspace to facilitate communication between researchers and context experts. The context experts were monetarily compensated for their work throughout the project.

\subsubsection{Hate Speech Definition}
There is no universally agreed-on definition for hate speech, and what counts as hate speech in one context may not be considered hate speech in another. This presents a challenge because accurate data annotation requires a standardized framework for consistency and reliability. We aimed to adopt a definition broad enough to potentially cover all instances of hate speech relevant to Myanmar, but was specific enough to avoid ambiguous interpretations by the annotators. To do this, we explored definitions provided by the UN, social media platform companies, and those widely used in hate speech research. 

According to the United Nations Strategy and Plan of Action on Hate Speech~\cite{guterres2019united}, hate speech is \textit{``any communication in speech, writing, or behavior, that attacks or uses pejorative or discriminatory language with reference to a person or a group on the basis of who they are, in other words, based on their religion, ethnicity, nationality, race, color, descent, gender or other identity factor"}. Researchers have developed definitions such as \textit{``language used to express hatred towards a targeted group or is intended to be derogatory, to humiliate, or to insult the members of the group."}~\cite{davidson2017automated} or \textit{``a deliberate attack directed towards a specific group of people motivated by aspects of the group’s identity"}~\cite{de2018hate}. 

Platforms often provide specific categories within which a person or group may be a target of hateful speech. For example, Facebook defines hate speech as \textit{``a direct attack against people — rather than concepts or institutions— on the basis of what we call protected characteristics: race, ethnicity, national origin, disability, religious affiliation, caste, sexual orientation, sex, gender identity and serious disease"}~\cite{facebook21} and Twitter defines it as \textit{``promote violence against or directly attack or threaten other people on the basis of race, ethnicity, national origin, caste, sexual orientation, gender, gender identity, religious affiliation, age, disability, or serious disease."}~\cite{twitter21}. Since Myanmar had not adopted a national definition for hate speech, we adapted a description for our annotation guideline composed of the UN's hate speech definition and included protected characteristics mentioned by Facebook. 
.
\subsubsection{Annotation Guidelines} 
Together with the context experts, we developed an annotation procedure designed to help standardize the annotation process. The annotation guideline for the task was made available in the English and Burmese languages. It contained our adapted definition of hate speech. We emphasized that hate speech must constitute dehumanizing or demeaning sentiment and be expressed because of who the target is based on some protected characteristics. We further added that hate speech may be directed towards an individual or a group \cite{elsherief2018hate} and could be explicit or implicit—the implicit case requiring an understanding of the context. We added a link to the original posts from the annotation files to enable this context inquiry. We highlighted instances when a speech is not regarded as hate speech, such as defamatory speech that does not invoke a protected characteristic or benign attacks against government policy. For example, a post saying \textit{``The Burmese military is corrupt!"} might be an attack on the military's integrity but does not constitute hate speech since the military is not a protected group. Annotators were provided with examples of posts that were hate speech and not hate speech, according to the guidelines. To validate our annotation guideline, we shared a copy with a team of legal experts based in Yangon, Myanmar, for feedback and edits. The experts provided three main feedback to improve the guideline.
\begin{enumerate}[i]
    \item The guideline should use an exhaustive list of protected characteristics instead of sampling of few examples that can leave too much room for annotator subjectivity.
    \item The guideline should clarify what action to take on hate speech regarding political holders. While political holders are protected from aforementioned attacks to protected characteristics, speech that attacks political decisions or ideology of the office holder is not hate speech.
    \item The annotator training should acknowledge the inherent challenges with defining an "attack" or a "demeaning post", especially for cases where hate speech is implicit. Since a post can be considered an attack or demeaning based on the perception of the recipient, there could be potential uncertainty in the labelled data.
\end{enumerate}
We edited our guidelines to address the concerns raised in (i) and (ii). For (iii), we relied on the competence of the context experts to reduce uncertainty. A final copy of our annotation guideline is provided in our Github repository.\footnote{\url{https://github.com/TID-Lab/myanmarhsc}}

\subsubsection{Data Collection}
Using Hatebase \cite{HateBase20}, an online repository of multilingual hate speech terms, we retrieved 128 crowdsourced Burmese hate terms. Three context experts manually inspected and removed 56 of the keywords considered overly context-sensitive and could potentially result in many benign posts. This inspection reduced the number of Burmese hate terms from Hatebase to 72. During the data collection phase, Phandeeyar~\cite{phandeeyar2021}, a technology organization based in Myanmar, released a lexicon of 88 hate terms resulting from their work tracking COVID-19 related hate speech on social media. We checked for possible appearances of the same words within the Hatebase and Phandeeyar lexicon sets. We found only two occurrences of an exact match. Another five were cases where a hate term in one set is the subset of another term in the other set. We combined both lexicons for a total of 158 hate terms for the next step. The list is also provided in our Github repository. \footnote{\url{https://github.com/TID-Lab/myanmarhsc}}

Next, we use CrowdTangle \cite{team2020crowdtangle}, a public insights service provided by Facebook which enables access to public groups and pages, to retrieve posts containing any of these select hate terms. Together with the context experts, we curated a dashboard of social and political pages in Myanmar using a combination of direct Facebook searches and local knowledge of popular social and political groups and pages. To identify new groups and pages, we snowball from already known pages to new pages that interact with them via shares or mentions. We downloaded historical Facebook posts from the CrowdTangle dashboard between October 2018 and June 2020 that contain any words from our hate lexicon. The downloaded data contained 43,996 posts. 

As a preprocessing step, we removed posts containing only URLs or blank shares of other posts. Next, we removed duplicate posts. This was a majority of the posts because CrowdTangle periodically provided updated interaction metrics for the same post leading to duplicates. We retained only the most recent versions of each post. We then randomized the posts in the dataset to remove any consecutive posts from the same account, group, or page. This process would help decrease anchoring bias in the annotations, which emerges when an annotator reads different posts from the same source and may become likely to label them similarly \cite{tversky1974judgment}. Finally, we removed posts with less than three syllables using the Myanmar word segmenter available as part of the Myanmar language tools package~\footnote{https://github.com/MyanmarOnlineAdvertising/myanmar\_language\_tools}. These posts will be too short and lack sufficient context for accurate labeling and feature engineering. After these preprocessing and filtering steps, 5,646 posts remained, which we proceeded to label.

\subsubsection{Labelling Plan}
Our recruited context experts understood Myanmar's socio-political terrain but were not necessarily hate speech experts. We were concerned that one remote training session might be insufficient to prepare them adequately, so we adopted a pairing strategy to boost annotator agreement. In this strategy, each annotator initially receives the same set of posts to label as one other annotator. Each annotator is asked to label the posts assigned to them independently. After labeling their initial batch of posts, they are then asked to schedule a call with their partner and the training facilitator to discuss their experience labeling the posts and address areas in which they disagreed. Each annotator was provided with a laptop and an internet connection to facilitate these conversations. There were eight annotators in our setup, resulting in four pairs. Each annotator received 100 posts per day over four days. After the fourth day, the remainder of the posts were divided equally amongst the annotators for individual labeling. Table~\ref{fig:pc} shows the number of posts labeled as Yes or No for each annotator, grouped with their paired annotator.

\begin{figure}[!ht]
    \center
    \includegraphics[width=\linewidth,scale = 0.5, trim=8 8 8 8,clip]{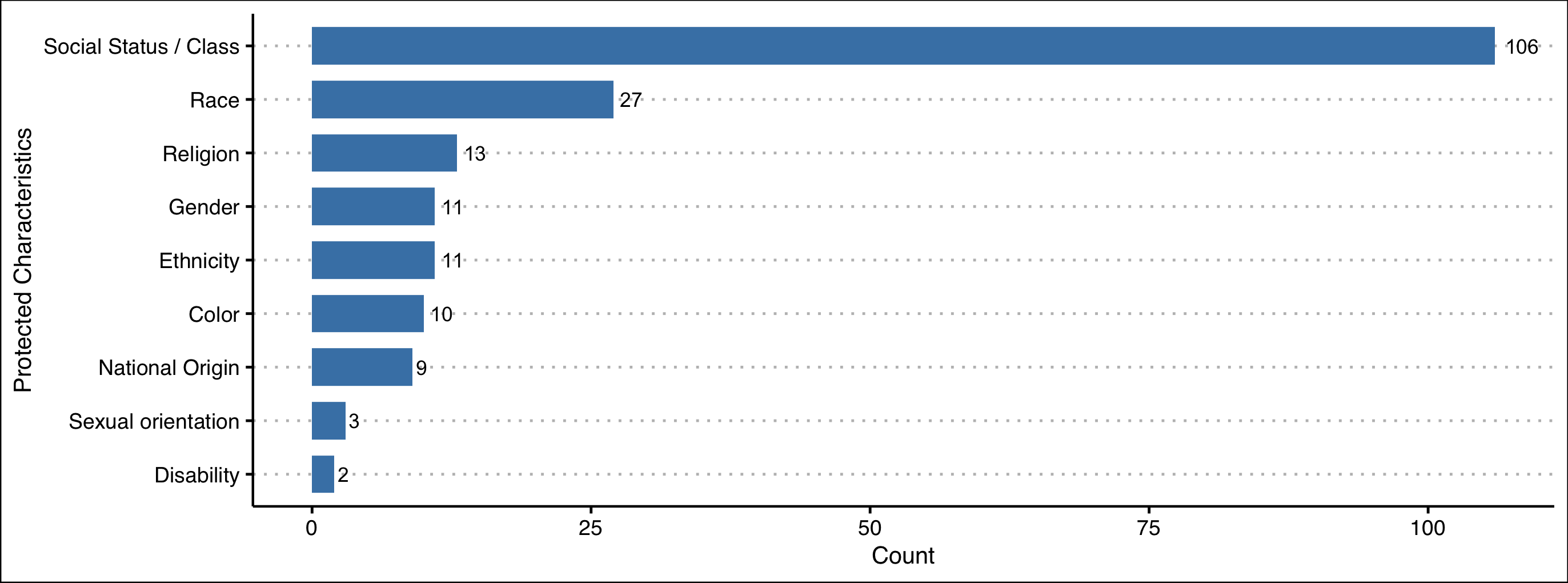}
    \caption{\label{schema}Protected characteristics invoked in posts labeled as hate speech}
\end{figure}

\subsubsection{Labelling Result}
After labeling, the final dataset contained 225 unique instances of posts labeled as hate speech. Figure \ref{fig:pc} shows the distribution of protected characteristics identified by the annotators as invoked in the post. We find that this is consistent with our expectations given Myanmar's socio-political scene. To measure the reliability of annotations, researchers rely on a measure of inter-annotator agreement to quantify the level of overlap among annotators on the labels they have chosen for each sentence. However, these scores such as Fleiss'~\cite{fleiss1971measuring} and Cohen's~\cite{cohen1960coefficient} kappas have been known to be affected by annotator bias and class imbalance. There is also no widely accepted kappa level for determining sufficient  reliability~\cite{ross2017measuring}. As a result, most hate speech research report very low agreement scores or do not report this value. For example, a batch of 100 posts given to one pair of annotators may include only one hate speech and 99 non-hate speech posts. If one annotator returns all 100 as not hate speech while the other accurately returns 99 not hate speech and one hate speech, their Cohen's kappa is 0.0 even though they agree 99\% of the time. Though percent agreement has been critiqued on the basis that it does not account for random agreements or guesses due to inadequate annotator training~\cite{mchugh2012interrater}, we adopt it as our measure of annotator improvement for our use case. We limit annotator guessing by training annotators extensively and setting up iterative peer feedback sessions at the end of each cycle.

\begin{figure}[h!]
    \centering
    \includegraphics[width=\linewidth,scale = 0.5, trim=8 8 8 8,clip]{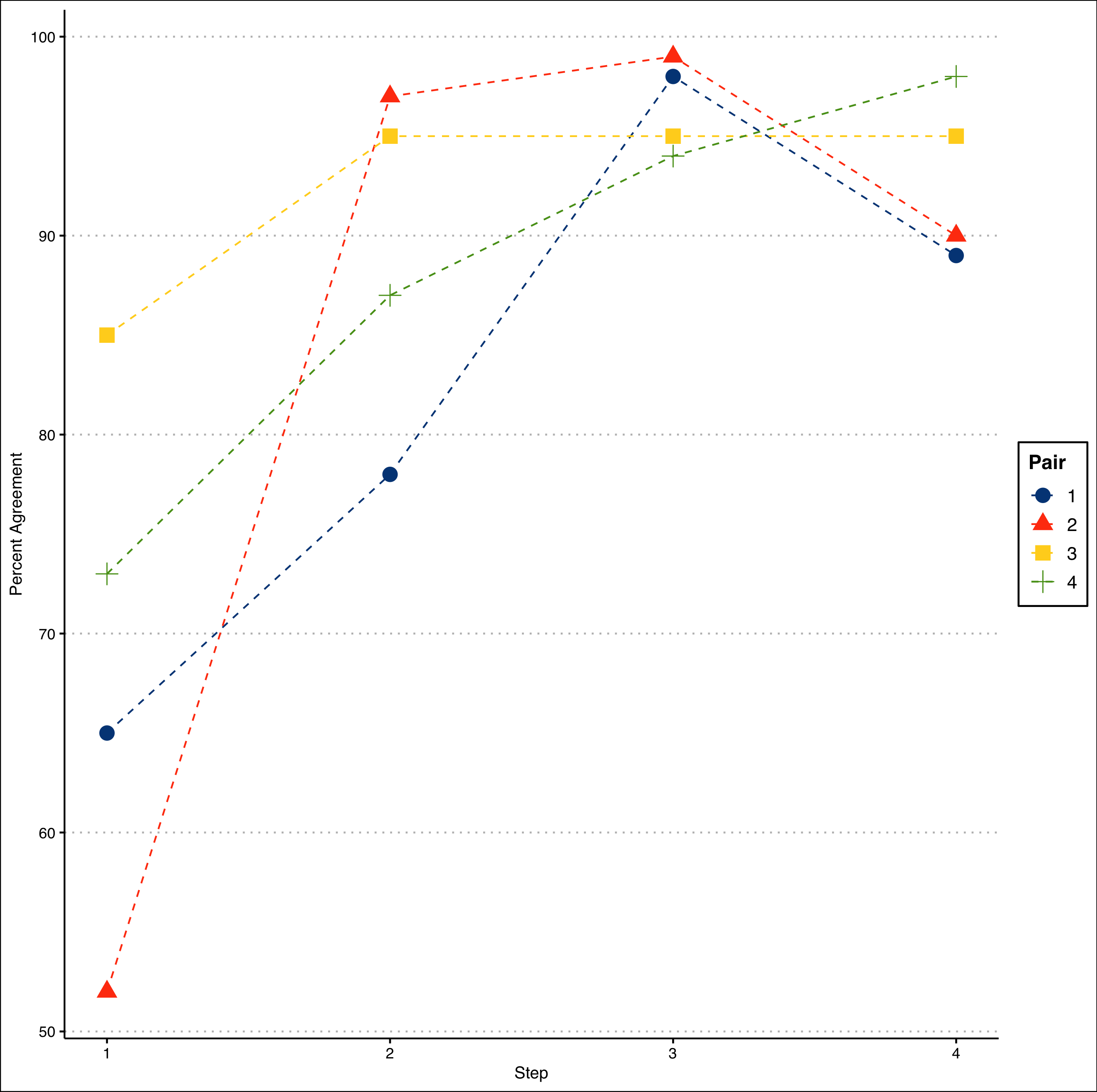}
    \caption{Percentage agreement within pairs of annotators for each training stage.}
    \label{fig:percag}
\end{figure}

We incorporated this peer learning model as part of the annotator training process to facilitate knowledge sharing where some annotators have some historical or current affairs context of some posts and build annotators' confidence from learning how much they are in sync with a peer. Figure~\ref{fig:percag} shows how percentage agreement among pairs of annotators increased after the first peer feedback session. Annotators were especially encouraged to discuss areas of disagreement. To avoid a situation where a pair is consistently wrong, we asked each pair to meet remotely with the training facilitator to discuss what they learned from that batch and confirm their results. By the fourth iteration, annotator agreement had increased from an average of 68.75\% to 93\%.

\begin{table*}[ht]
\centering
\caption{Number of posts labeled as Yes or No  by each annotator.}
\label{tab:percscore}
\begin{tabular}{|*{17}{c|}}
\hline
\multirow{3}*{\textbf{Step}} &
\multicolumn{4}{|c}{\textbf{Pair 1}} & \multicolumn{4}{|c}{\textbf{Pair 2}} & \multicolumn{4}{|c}{\textbf{Pair 3}} & \multicolumn{4}{|c|}{\textbf{Pair 4}} \\ \cline{2-17} &
\multicolumn{2}{|c}{\textbf{A1}} & \multicolumn{2}{|c}{\textbf{A2}} & \multicolumn{2}{|c}{\textbf{A3}} & \multicolumn{2}{|c|}{\textbf{A4}} & \multicolumn{2}{|c}{\textbf{A5}} & \multicolumn{2}{|c}{\textbf{A6}} & \multicolumn{2}{|c}{\textbf{A7}} & \multicolumn{2}{|c|}{\textbf{A8}} \\ \cline{2-17} &
Yes & No & Yes & No & Yes & No & Yes & No & Yes & No & Yes & No & Yes & No & Yes & No \\ \hline
1 & 60 & 40 & 76 & 24 & 59 & 41 & 13 & 87 & 10 & 90 & 17 & 83 & 17 & 83 & 34 & 66 \\ \hline
2 & 19 & 81 & 4 & 96 & 3 & 97 & 2 & 98 & 5 & 95 & 2 & 98 & 0 & 100 & 13 & 87 \\ \hline
3 & 2 & 98 & 4 & 96 & 1 & 99 & 1 & 99 & 4 & 96 & 1 & 99 & 5 & 95 & 1 & 99 \\ \hline
4 & 12 & 88 & 4 & 96 & 4 & 96 & 6 & 96 & 9 & 91 & 6 & 94 & 2 & 98 & 0 & 100 \\ \hline
\end{tabular}
\end{table*}

\subsubsection{Text classification} 
We designed a four-stage data preprocessing pipeline: first, we converted the Zawgyi character to Unicode. We used an open-source Myanmar language tools library~\cite{khine2020tools} for this step. Next, we removed posts predominantly in languages that are not Burmese. We removed emojis. We retained instances of code-switching between English and Burmese. In addition, since the Burmese language does not consistently use white spaces to mark word boundaries, we used the Myanmar language tools~\cite{khine2020tools} library to perform word-level segmentation on the text splitting each post into tokens. Finally, we used an openly available Burmese stop words list~\cite{maung2008rule} to remove stop words from each remaining post in our dataset.

\subsubsection{Classification Models}
We employed classical machine learning classification algorithms (Support Vector Machines~\cite{cortes1995support}, Balanced Random Forests~\cite{chen2004using}, and FastText~\cite{bojanowski2017enriching}) to conform to similar constraints in contexts where models such as pre-trained Transformers are unavailable. We ran several feature combinations of n-grams, term-frequency weighting, and hyper-parameter searches to select the best combination. We show precision, recall, and F-1 scores for the models in Table \ref{tab:modelperf}. From Table \ref{tab:modelperf}, we observe that the FastText model (with oversampling) performed best across all three metrics.

\begin{table}[h]
  \caption{Model performance for hate speech classification}
  \label{tab:modelperf}
  \begin{tabular}{lrrr}
    \toprule
    Model & Precision & Recall & F1-score\\
    \midrule
    SVM & 0.48 & 0.50 & 0.49\\
    SVM (with oversampling) & 0.88 & 0.87 & 0.87\\
    BRF  & 0.53 & 0.70 & 0.45\\
    BRF (with oversampling)  & 0.88 & 0.87 & 0.87\\
    FastText & 0.84 & 0.63 & 0.69\\
    FastText (with oversampling) & 0.93 & 0.92 & 0.92\\
  \bottomrule
\end{tabular}
\end{table}

\subsection{Context Experts as Model Validators}
Beyond the cross-validation techniques discussed in the previous section, we sought to validate our model qualitatively to understand which cases led to model error and why. There are limited tools for performing these kinds of model validation~\cite{gao2019ai}. For our analysis, we collected live data from Facebook different from those in our training and tests set and passed these data for inference on our best-performing model. We provided a sample of the data and asked our context experts to label them. We then show the model decisions to the context experts and discuss areas where they disagreed with the model. 

We identified two major error types from the model from this step. First, false-positive cases where benign posts containing terms in the hate lexicons or words typically used in hate speech posts and highly represented in the training data were wrongly flagged as hate speech by the model. Second, false-negative cases were hateful posts that did not contain the archetypal hate terms and were not classified as hate speech by the model.

%% file: sections/findings.tex
In this section, we discuss our findings from the collaborative model with context experts, which we have described in the previous section. While we drew most of these observations from our process in Myanmar, they demonstrate similar issues in other low-resource language settings where machine learning may be applied. An appreciation of these findings is crucial for designing efficient machine learning systems. We have identified three central issues: (i) exploring avenues to boost ground truth data, (ii) incentivizing data work for machine learning, and (iii) engendering open data-sharing practices.

\subsection{Exploring avenues to boost ground truth data}
A major challenge with hate speech detection tasks in low-resource language settings is limited training data in target languages. This limitation is often due to the rarity of hate speech in proportion to the amount of available social media data and the infeasibility of labeling the entire dataset on a given platform~\cite{madukwe2020data}. Much of the progress in the field of machine learning has been driven by the availability of benchmark datasets such as ImageNet \cite{deng2009imagenet} for computer vision, and GLUE \cite{wang2018glue} for natural language processing. While such datasets have sprung up for hate speech detection in the English language \cite{waseem2016hateful, mathew2020hatexplain}, researchers often filter out data not in English, leaving more work to do for very low-resource languages. This lack of training data can hamper machine learning research within these contexts. The ML4D community can adopt the process described in this work to bootstrap the development of hate speech training datasets in diverse contexts.

A caveat to note is that engaging in elaborate data collection and labeling efforts do not always guarantee significant outcomes. We found that only less than 4\% of the entire Facebook posts in our dataset were labeled by annotators as hate speech. To boost training data, practitioners have relied on curated hate speech lexicons in the target languages. One example is the PeaceTech Lab Lexicons, a series of hate speech terms explaining inflammatory social media keywords and offering counter-speech suggestions to combat the spread of hate speech~\cite{peacetechlab}. The PeaceTech Lab has curated hate lexicons for languages in conflict-affected countries such as the Democratic Republic of the Congo, Sudan, and Lebanon. Though keyword approaches are ineffective when used alone, they can help researchers select a subset of data to work with. Organizing civil society workshops or other avenues for crowdsourcing hate terms can be beneficial for supporting hate speech detection work.

Our findings reveal how quickly new hate terms emerge on social media. This dynamism is primarily due to changes in political and social concerns. For instance, online actors may derive new hate terms as social media discourse changes from talking about a local election to focusing on a pandemic~\cite{ziems2020racism}. Context experts can help identify when social media topics drift or when users find creative ways to guise hate speech. 

\subsection{Incentivizing data work for machine learning}
Data work is ``any human activity related to creating, collecting, managing, curating, analyzing, interpreting, and communicating data''~\cite{bossen2019data}. This fundamental component of machine learning is often undervalued in research and practice~\cite{wagstaff2012machine} and can lead to negative outcomes because data is critical for effective machine learning systems~\cite{polyzotis2018data}. Sambasivan et al.~\cite{sambasivan2021everyone} discuss the challenges with data cascades in high-stakes AI, which they define as compounding events over time that result from the undervaluing of data quality by researchers and practitioners. These data cascades often have significant effects in low-resource language settings due to the lack of existing ground data and supporting infrastructure for data collection and processing. We identified some of the cascade factors identified by Sambasivan et al.~\cite{sambasivan2021everyone} in our work, namely, incentives in AI and data education. First, we observed that while stakeholders and partners might identify with and value the role of AI in addressing the problem of low-resource hate speech, they often did not place similar consideration on the invisible and challenging data work. Second, context experts and partners lack experience in creating the kinds of complex and quality datasets that hate speech research requires. Stakeholders ought to see the value in data work to appreciate the time spent on training context experts.

At the onset of communications with sponsors and partners, it is crucial to emphasize the importance of data work for machine learning and the vital role that context experts will play. This step may imply a more costly engagement with context experts, both in time and financial compensation. We learned that a comprehensive data training plan is helpful at the start of the project to address the data education challenge. 

\subsection{Supporting open data sharing practices}
Social media researchers in low-resource contexts do not often have the freedom to choose what platforms to work on. This decision is mainly driven by which platform is widely used and most likely to have relevant data within a given context. With over 21 million active users in 2019~\cite{kemp19}, our work in Myanmar primarily focused on Facebook. Like any other platform, Facebook offers unique affordances and constraints for data access and sharing. For example, the company grants privileged data access to researchers via its CrowdTangle API service. Yet, its terms of use do not allow researchers to share any data with people outside CrowdTangle. Such practices make it challenging to freely collaborate with other teams who may also have access to data but are bound by the platform's terms.

These limitations posed by platforms motivate community-based data-sharing workflows led by context experts and built on trust, ethics, and shared values. Data sharing has been discussed as an effective means for scientific progress, especially within developing countries \cite{chawinga2019global, sebake2012assessing}. However, there are lots of structural, organizational, cultural, and ethical complexities that undermine data sharing~\cite{anane2018you}. Exploring data-sharing practices that actively address power imbalances, understands potential benefits and risks, and comply with local norms and cultures~\cite{abebe2021narratives} will help produce a critical mass of relevant data that can be useful for tackling low-resource hate speech.

Furthermore, significant aspects of a country's history are often undocumented digitally, only existing as paper documents within locked-out cabinets or as informal knowledge passed on from one generation to another. In some cases, only a fraction of the context experts know the historical and cultural issues that underlie hateful speech posted online. Our annotation model addresses this knowledge imbalance with the peer feedback model discussed as part of our process. 

%% file: sections/conclusion.tex
There is an urgent need to identify ways to tackle hate speech on social media in low-resource language settings. We have argued that addressing this problem will require community-based approaches that combine a deep understanding of social and political contexts with automated tools to process the vast amount of content produced online. In this paper, we presented findings from our remote study on the automatic detection of hate speech on Facebook in Myanmar. We have developed a systematic process for collaborating with context experts covering critical stages of data collection, annotation, and model validation strategies. Our work offers insights for researchers and practitioners in machine learning for development (ML4D) and highlights challenges stemming from small and imbalanced datasets, the need to balance non-glamorous data work and stakeholder priorities, and closed data sharing practices. These findings motivate further research exploring strategies for data augmentation, non-text-based detection models, multimodal hate speech detection, and best practices for working with non-machine learning experts on machine learning-powered projects. 

\textbf{Ethics statement:} 
We hope this work can support efforts to protect the integrity of social media platforms and defend democracies against the threats of hate speech, disinformation, and misinformation. We understand that automated tools such as those proposed in this work could be misused to target and oppress dissenting voices, especially within authoritarian regimes. We unequivocally state that such use will be contrary to our core goal of offering ideas for combating harmful content online.

%% file: sections/future.tex
We now outline some promising directions for future research work in this area addressing issues of limited ground truth data, non-text-based techniques, multimodality, and non-expert collaboration:
 
\subsection{Data augmentation for low-resource hate speech detection} 
Data augmentation involves strategies for increasing training examples for machine learning tasks without explicitly collecting new data. Data augmentation has received recent attention in natural language processing research due to increased work in new domains with limited data and the need for substantial amounts of training data for large neural networks~\cite{feng2021survey}. As our findings have shown, running entire data labeling schemes may not necessarily lead to more data, especially for tasks where true labels are scarce. Low-resource hate speech detection work can benefit from data augmentation strategies that generate new data to augment the sparsity in new contexts. This technique can take the form of generating entirely new corpora from validated ground truths in high-resource settings or creating new variations of existing small data in the target language.
 
\subsection{Exploring network-based modeling approaches} 
Current hate speech detection tools in low-resource language settings mostly rely on text-based user-generated data and less on other contextual information such as linked news sources, parallel comments from local news websites, platform metadata, and other social network data. Incorporating these additional contexts into existing models could improve single-instance hate speech detection and coordinated hate attacks that can be difficult to detect in real time. One idea is to focus on understanding problematic actors within these contexts and identifying relationships that increase the likelihood of spreading hateful content. Some recent works have explored graph-based models for this task~\cite{del2019you, mishra2018author, beatty2020graph} but little is known about how these methods translate to low-resource contexts. A practical solution could help address the challenges posed by limited training data and other vulnerabilities of text-based methods such as topic drift and adversarial attacks.

\subsection{Multimodal hate speech detection} 
A vast amount of hate speech posts in low-resource language settings are multimodal, often containing a combination of images, audio content, videos, live streams, external links, etc. Models trained with data on a single modality might be insufficient to address the problem effectively. Presently, content moderators have to watch hours of videos in their native language to identify hateful rhetoric embedded in the videos before recommending take-downs. This process is challenging to scale, takes lots of moderation hours, and videos tend to spread faster than platform action. Image-text datasets such as MMHS150K~\cite{gomez2020exploring} and the Hateful Memes~\cite{kiela2020hateful} have been released in English for this task. Further work is needed for hate speech detection for multimodal content for low-resource contexts.

\subsection{Active learning and uncertainty-aware predictions}
Given established concerns over the scarcity of true labels and the prevalence of noisy data in the domain of hate speech detection, an active learning strategy might help boost model performance in practice. The idea is that a machine learning model can perform better with fewer labeled training instances if it can choose a reasonable sample to learn from~\cite{settles.tr09}. This is especially useful for systems that are deployed in the wild where the machine learning model can provide strategic queries to human annotators and retrain on a valuable set of new training data. An uncertainty-aware sampling strategy can help ensure that the uncertainty resulting from the data imbalance is well-calibrated~\cite{lewis1994sequential}.

\subsection{Working with non-experts on machine learning deployment projects}  
Machine learning work with non-experts requires a degree of care to avoid the trap of participation-washing where their involvement is either outrightly performative or only aimed at extracting free labor and consultation~\cite{sloan2020participation}. We are only beginning to scratch the surface of what forms effective AI collaboration with context experts may take. In this work, we have shown how we defined and scoped the problem with context experts, collected and labeled the data, and validated the model. Nonetheless, the technical gaps in the context experts' knowledge of machine learning implied that some parts of the modeling process were inadvertently black-boxed from them. This gap may present challenges for machine learning project sustainability, and further work is needed to understand ways to identify and mitigate the critical (people, process, and cultural~\cite{kumar2006impact}) failure factors that might hinder the long-term success of AI deployments in low-resource contexts.